\documentclass[useAMS,usenatbib]{mn2e}
\usepackage{graphicx}
\usepackage{epsfig}

\title[Progenitors of SNe Ia]{Helium star donor channel for the progenitors of type Ia supernovae}
\author[B. Wang, X. Meng, X. Chen and Z. Han]
{B. Wang$^{\rm 1,2}$ \thanks{E-mail: wangbo@ynao.ac.cn}, X.
Meng$^{\rm 3}$, X. Chen$^{\rm 1}$ and Z. Han $^{\rm 1}$  \\
$^1$ National Astronomical Observatories/Yunnan Observatory, the
Chinese Academy of Sciences, Kunming 650011, China\\
$^2$ Graduate School of the Chinese Academy of Sciences, Beijing
100049, China\\
$^3$ Department of Physics and Chemistry, Henan Polytechnic
University, Jiaozuo 454003, China}

\begin{document}
\pagerange{\pageref{firstpage}--\pageref{lastpage}} \pubyear{2008}
\maketitle

\label{firstpage}

\begin{abstract}
Type Ia supernovae (SNe Ia) play an important role in astrophysics,
especially in the study of cosmic evolution. There are several
progenitor models for SNe Ia proposed in the past years. In this
paper, we have carried out a detailed study of the He star donor
channel, in which a carbon-oxygen white dwarf (CO WD) accretes
material from a He main sequence star or a He subgiant to increase
its mass to the Chandrasekhar mass. Employing Eggleton's stellar
evolution code with an optically thick wind assumption, and adopting
the prescription of Kato \& Hachisu (2004) for the mass accumulation
efficiency of the He-shell flashes onto the WDs, we performed binary
evolution calculations for about 2600 close WD binary systems.
According to these calculations, we mapped out the initial
parameters for SNe Ia in the orbital period--secondary mass ($\log
P^{\rm i}-M^{\rm i}_2$) plane for various WD masses from this
channel. The study shows that the He star donor channel is
noteworthy for producing SNe Ia (i.e. $\sim 1.2\times10^{-3}\,{\rm
yr}^{-1}$ in the Galaxy), and that the progenitors from this channel
may appear as supersoft X-ray sources. Importantly, this channel can
explain SNe Ia with short delay times ($\la 10^{8}$\,yr), which is
consistent with recent observational implications of young
populations of SN Ia progenitors.


\end{abstract}


\section{INTRODUCTION}\label{INTRODUCTION}
Type Ia Supernovae (SNe Ia) are excellent cosmological distance
indicators due to their high luminosities and remarkable uniformity.
They have been applied successfully in determining cosmological
parameters (e.g. $\Omega$ and $\Lambda$; Riess et al. 1998;
Perlmutter et al. 1999). It is widely believed that SNe Ia are
thermonuclear explosions of carbon-oxygen white dwarfs (CO WDs)
accreting matter from their companions (see a review of Nomoto et
al. 1997). However, several key issues related to the nature of
their progenitor systems and the physics of the explosion mechanisms
are still not well understood (Hillebrandt \& Niemeyer 2000;
R\"{o}pke \& Hillebrandt 2005; Wang et al. 2008; Podsiadlowski et
al. 2008), which may raise doubts about the distance calibration
being purely empirical and on the basis of the SN Ia sample of the
low red-shift Universe.



At present, two SN Ia explosion models are discussed frequently,
that is, Chandrasekhar (Ch) mass model and sub-Chandrasekhar
(sub-Ch) mass model. The synthetic spectra of the Ch mass model are
in excellent agreement with the early time spectra of most SNe Ia,
while that of the sub-Ch mass model has difficulty in matching
observations (H\"{o}flich $\&$ Khokhlov 1996). A CO WD can increase
its mass to the Ch mass through a single degenerate (SD) scenario,
where the CO WD accretes H/He-rich material from a non-degenerate
companion star (Whelan \& Iben 1973; Nomoto et al. 1984), or through
a double degenerate (DD) scenario, where another CO WD mergers with
it and the total mass of the two CO WDs is larger than the Ch mass
limit (Iben \& Tutukov 1984; Webbink 1984). Theoretically, it is
suggested that the DD scenario likely leads to an accretion-induced
collapse rather than a SN Ia (Nomoto \& Iben 1985; Saio \& Nomoto
1985; Timmes et al. 1994).




For the SD Ch scenario, the companion is probably a MS star or a
slightly evolved subgiant star (WD + MS channel), or a redgiant star
(WD + RG channel) (Hachisu et al. 1996, 1999a, 1999b; Li $\&$ van
den Heuvel 1997; Langer et al. 2000; Han $\&$ Podsiadlowski 2004,
2006; Chen \& Li 2007; Meng et al. 2008; Han 2008). Meanwhile, a CO
WD may also accrete material from a He star companion to increase
its mass to the Ch mass, which is known as the He star donor channel
in this paper. The study of Iben $\&$ Tutukov (1994) showed that WD
+ He star systems can be formed via binary evolutions. Yoon $\&$
Langer (2003) have carried out the evolution of a CO WD + He star
system with a $1.0\,M_{\odot}$ CO WD and a $1.6\,M_{\odot}$ He star
in a 0.124 day orbit. In this binary, the WD accretes He from the He
star and grows in mass to the Ch mass. It is believed that WD + He
star systems are generally originated from intermediate-mass binary
systems. Thus, this channel may explain SNe Ia with short delay
times (Mannucci et al. 2006; Aubourg et al. 2008). At present,
however, there are no comprehensive investigations on this channel.

The existence of WD + He star systems is supported by some
observations. For example, KPD 1930+2752 is suggested to be such a
system (Maxted et al. 2000; Geier et al. 2007). Another example is
from the discovery of a He nova, V455 Pup (Ashok \& Banerjee 2003;
Kato \& Hachisu 2003). Recently, Kato et al. (2008) presented a
free-free emission dominated light curve model of V445 Pup, based on
an optically thick wind theory (Kato \& Hachisu 1994; Hachisu et al.
1996). The light curve fitting in their study showed that the mass
of the WD is lager than $1.35\,M_{\odot}$, and half of the accreted
matter remains on the WD, leading to the mass increase of the WD.
Thus, V445 Pup is suggested to be a strong candidate of SN Ia
progenitor (Kato et al. 2008).
%

The purpose of this paper is to study the He star donor channel for
SN Ia progenitors comprehensively and systemically, and then to
determine the valid parameter spaces for SNe Ia, which can be used
in binary population synthesis (BPS) studies. In Section 2, we
describe the numerical code for the binary evolution calculations
and the grid of the binary models. The binary evolutionary results
is shown in Section 3. In Section 4, we estimate SN Ia birthrate of
this channel. Finally, a discussion is  given in Section 5, and a
summary in Section 6.





\section{BINARY EVOLUTION CALCULATIONS}\label{BINARY EVOLUTION CALCULATIONS}
In WD + He star systems, the He star fills its Roche lobe at He MS
or He subgiant stage, and then the mass transfer begins. The He star
transfers some of its material onto the surface of the WD, which
increases the mass of the WD as a consequence. If the WD grows to
1.378\,$M_{\odot}$, we assume that the WD explodes as a SN Ia. Here
we use Eggleton's stellar evolution code (Eggleton 1971, 1972, 1973)
to calculate the binary evolutions of WD + He star systems. The code
has been updated with the latest input physics over the last three
decades (Han et al. 1994; Pols et al. 1995, 1998). Roche lobe
overflow (RLOF) is treated within the code described by Han et al.
(2000). We set the ratio of mixing length to local pressure scale
height, $\alpha=l/H_{\rm p}$, to be 2.0. The opacity tables are
compiled by Chen \& Tout (2007) from Iglesias \& Rogers (1996) and
Alexander \& Ferguson (1994). In our calculations, the He star
models are composed of He abundance $Y=0.98$ and metallicity
$Z=0.02$, and all calculations for the He stars are carried out
without enhanced mixing (i.e. the overshooting parameter,
$\delta_{\rm ov}$, is taken to be zero; see Dewi et al. 2002). In
addition, orbital angular momentum loss due to gravitational wave
radiation (GWR) is included by adopting a standard formula presented
by Landau \& Lifshitz (1971),
\begin{equation}
{d\,\ln J_{\rm GR}\over dt} = -{32G^3\over 5c^5}\,{M_{\rm WD} M_2
(M_{\rm WD}+M_2)\over a^4},
\end{equation}
where $G$, $c$, $M_{\rm WD}$ and $M_2$ are the gravitational
constant, vacuum speed of light, the mass of the accreting WD and
the mass of the companion He star, respectively.

Instead of solving stellar structure equations of a WD, we use an
optically thick wind model (Kato \& Hachisu 1994; Hachisu et al.
1996) and adopt the prescription of Kato \& Hachisu (2004) (KH04)
for the mass accumulation efficiency of He-shell flashes onto the
WD. If the mass transfer rate, $|\dot M_2|$, is above a critical
rate, $\dot M_{\rm cr}$, we assume that He burns steadily on the
surface of the WD and that the He-rich matter is converted into C
and O at a rate $\dot M_{\rm cr}$. The unprocessed matter is lost
from the system, presumably in the form of an optically thick wind
at a mass loss rate $\dot M_{\rm wind}=|\dot M_2| - \dot M_{\rm
cr}$. The critical mass transfer rate is
\begin{equation}
\dot M_{\rm cr}=7.2\times 10^{-6}\,(M_{\rm
WD}/M_{\odot}-0.6)\,M_{\odot}\,\rm yr^{-1},
\end{equation}
based on WD models computed with constant mass accretion rates
(Nomoto 1982).

Following assumptions are adopted when $|\dot M_2|$ is smaller than
$\dot M_{\rm cr}$. (1) If $|\dot M_2|$ is less than $\dot M_{\rm
cr}$ but higher than the minimum accretion rate of stable He-shell
burning, $\dot M_{\rm st}$ (KH04), it is assumed that the He-shell
burning is stable and that there is no mass loss. (2) If $|\dot
M_2|$ is less than $\dot M_{\rm st}$ but higher than the minimum
accretion rate of weak He-shell flashes, $\dot M_{\rm
low}=4.0\times10^{-8}\,M_{\odot}\,\rm yr^{-1}$ (Woosley et al.
1986), He-shell flashes occur and a part of the envelope mass is
assumed to be blown off from the surface of the WD. The mass growth
rate of WDs in this case is linearly interpolated from a grid
computed by KH04, where a wide range of WD mass and accretion rate
were calculated in the He-shell flashes. (3) If $|\dot M_2|$ is
lower than $\dot M_{\rm low}$, the He-shell flashes will be so
strong that no mass can be accumulated onto the WD.

We define the mass growth rate of the CO WD, $\dot{M}_{\rm CO}$, as
 \begin{equation}
 \dot{M}_{\rm CO}=\eta _{\rm He}|\dot{M}_{\rm 2}|,
  \end{equation}
where $\eta _{\rm He}$ is the mass accumulation efficiency for
He-shell burning. According to the assumptions above, the values of
$\eta _{\rm He}$ are:
\begin{equation}
\eta_{\rm He}= \left\{ \begin{array}{l@{\quad,\quad}l}
\dot M_{\rm cr}\over |\dot M_2| &  |\dot{M_2}|>\dot{M}_{\rm cr},\strut\\
1\, &  \dot{M}_{\rm cr}\ge |\dot{M_2}|\ge \dot{M}_{\rm st},\strut\\
\eta'_{\rm He}\, &  \dot{M}_{\rm st}> |\dot{M_2}|\ge \dot{M}_{\rm low},\strut\\
0 &  |\dot{M_2}|< \dot{M}_{\rm low}.\strut\\
\end{array} \right.
\end{equation}


We incorporate the prescriptions above into Eggleton's stellar
evolution code and follow the binary evolutions of WD + He star
systems. The mass lost from these systems is assumed to take away
specific orbital angular momentum of the accreting WD. We have
calculated about 2600 WD + He star systems, and obtained a large,
dense model grid. The initial mass of the donor stars, $M_{\rm
2}^{\rm i}$, ranges from $0.85\,M_{\odot}$ to $3.1\,M_{\odot}$; the
initial mass of the CO WDs, $M_{\rm WD}^{\rm i}$, is from
$0.865\,M_{\odot}$ to $1.20\,M_{\odot}$; the initial orbital period
of the binary systems, $P^{\rm i}$, changes from the minimum value,
at which a He zero-age MS (He ZAMS) star would fill its Roche lobe,
to $\sim 316$ days, where the He star fills its Roche lobe at the
end of the Hertzsprung gap.


\section{BINARY EVOLUTION RESULTS}\label{sect:BINARY EVOLUTION RESULTS}
\subsection{Typical binary evolution calculations}

\begin{figure*}
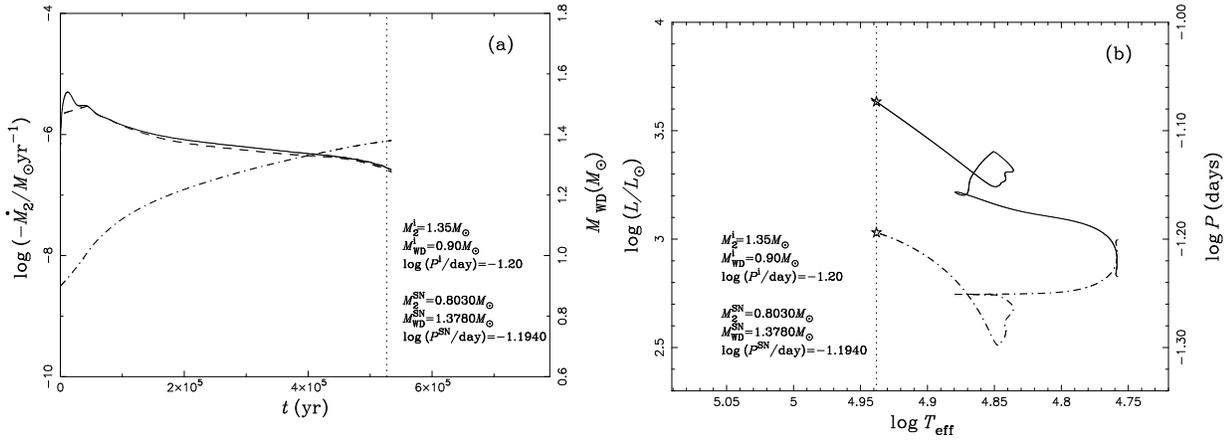

\centerline{\epsfig{file=mdot1.ps,angle=270,width=8cm}\ \
\epsfig{file=hrd1.ps,angle=270,width=8cm}} \caption{A representative
case of binary evolution calculations, in which the binary system is
in the weak He-shell flashes phase at the moment of the SN
explosion. In panel (a), the solid, dashed and dash-dotted curves
show $\dot M_2$, $\dot M_{\rm CO}$ and $M_{\rm WD}$ varing with
time, respectively. In panel (b), the evolutionary track of the
donor star is shown as a solid curve and the evolution of orbital
period is shown as a dash-dotted curve. Dotted vertical lines in
both panels and asterisks in panel (b) indicate the position where
the WD is expected to explode as a SN Ia. The initial binary
parameters and the parameters at the moment of the SN Ia explosion
are also given in these two panels.}
\end{figure*}

\begin{figure*}
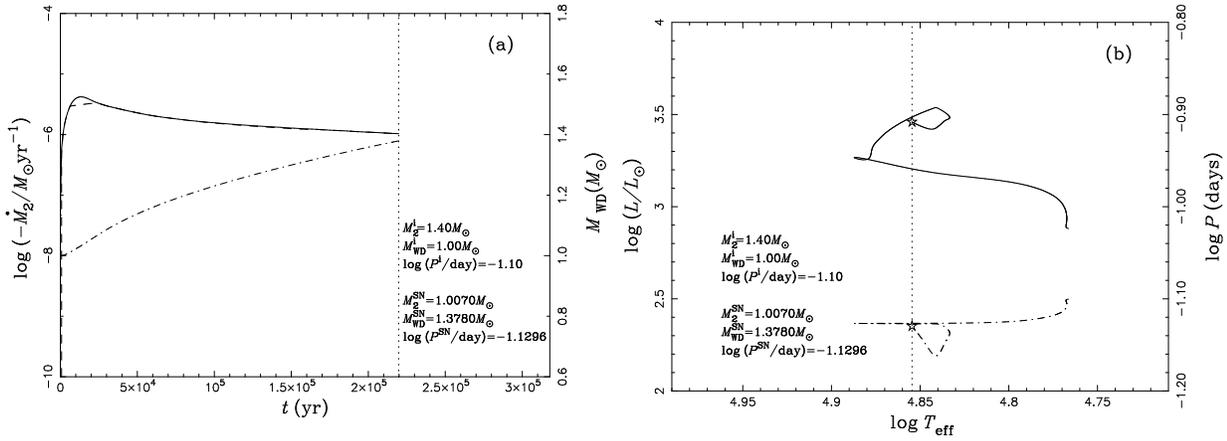

\centerline{\epsfig{file=mdot2.ps,angle=270,width=8cm}\ \
\epsfig{file=hrd2.ps,angle=270,width=8cm}} \caption{Similar to
Fig.1, but the binary system is in the stable He-shell burning phase
at the moment of the SN explosion.}
\end{figure*}

\begin{figure*}
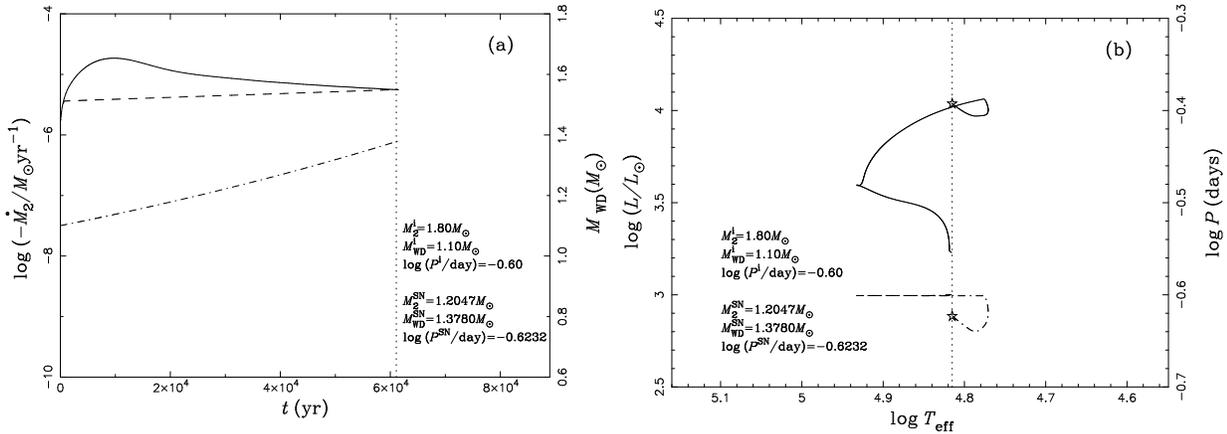

\centerline{\epsfig{file=mdot3.ps,angle=270,width=8cm}\ \
\epsfig{file=hrd3.ps,angle=270,width=8cm}}\caption{Similar to Fig.1,
but the binary system is in the optically thick wind phase at the
moment of the SN explosion.}
\end{figure*}

\begin{figure*}
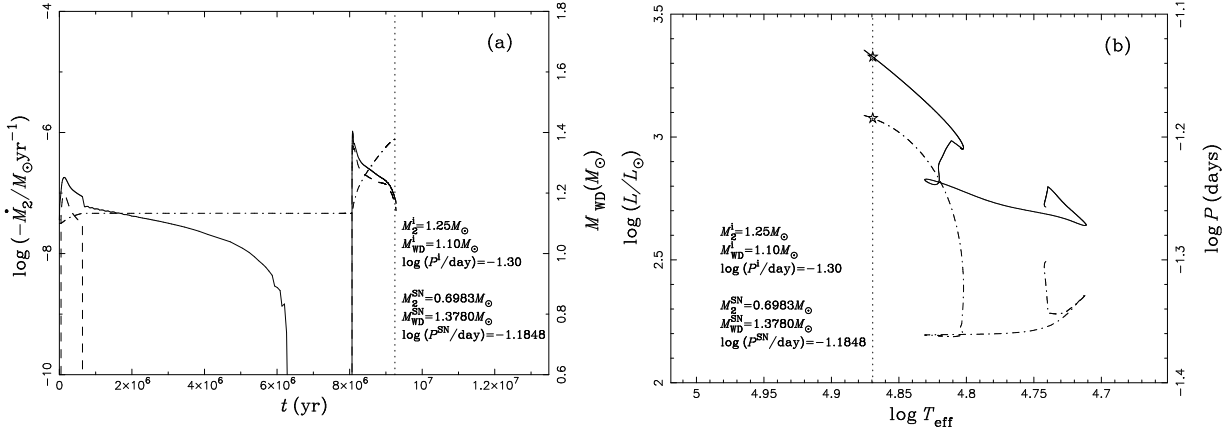

\centerline{\epsfig{file=mdot4.ps,angle=270,width=8cm}\ \
\epsfig{file=hrd4.ps,angle=270,width=8cm}}\caption{Similar to Fig.1,
but this case differs from the three above. Before the SN explosion,
the binary system experiences two mass transfer phases.}
\end{figure*}

In Figs 1-4, we present three representative cases in our binary
evolution calculations according to the condition of the binary
system at the moment of SN explosion, and display one special case
for producing SNe Ia. In panels (a) of these figures, we show the
$\dot M_2$, $\dot M_{\rm CO}$ and $M_{\rm WD}$ varing with time, and
the panels (b) are the evolutionary tracks of the He stars in the
Hertzsprung-Russell diagrams, where the evolutions of the orbital
periods are also shown.


(i) {\em Case 1} (see Fig. 1): The binary system is in the weak
He-shell flash phase at the moment of the SN explosion. The binary
shown in this case is ($M_2^{\rm i}$, $M_{\rm WD}^{\rm i}$, $\log
(P^{\rm i}/{\rm day})$) $=$ (1.35, 0.9, -1.20), where $M_2^{\rm i}$,
$M_{\rm WD}^{\rm i}$ and $P^{\rm i}$ are the initial mass of the He
star and of the CO WD in solar masses, and the initial orbital
period in days, respectively. In this case, the He star fills its
Roche lobe after the exhaustion of central He (it now contains a CO
core), and this results in Case BB mass transfer \footnote{We
distinguish case BB (Roche lobe overflow occurs after He-core
burning but before carbon ignition) from case BA (in which mass
transfer is initiated during He-core burning) (see Dewi et al.
2002).}. The mass transfer rate $|\dot{M}_{\rm 2}|$ exceeds $\dot
M_{\rm cr}$ soon after the onset of RLOF, leading to a wind phase,
where a part of the transferred mass is blown off in an optically
thick wind, and the left is accumulated onto the WD. After about
$5\times10^{4}$\,yr, $|\dot{M}_{\rm 2}|$ drops below $\dot M_{\rm
cr}$ but still higher than $\dot M_{\rm st}$. Thus, the optically
thick wind stops and the He-shell burning is stable. With the
continuous decreasing of $|\dot{M}_{\rm 2}|$, the system enters into
a weak He-shell flash phase after about $5\times10^{4}$\,yr. The WD
always grows in mass until it explodes as a SN Ia in the weak
He-shell flash phase. At the SN explosion moment, the mass of the
donor star is $M^{\rm SN}_2=0.8030\,M_{\odot}$ and the orbital
period $\log (P^{\rm SN}/{\rm day})=-1.1940$.



(ii) {\em Case 2} (see Fig. 2): The binary system is in the stable
He-shell burning phase at the moment of the SN explosion. The binary
in this case is ($M_2^{\rm i}$, $M_{\rm WD}^{\rm i}$, $\log (P^{\rm
i}/{\rm day})$) $=$ (1.40, 1.00, -1.10). The binary evolves in a
similar way to that of {\em Case 1}, but it is in stable He-shell
burning stage when the CO WD reaches $1.378\,M_{\odot}$. The binary
parameters at $M_{\rm WD}=1.378\,M_{\odot}$ are $M^{\rm
SN}_2=1.0070\,M_{\odot}$ and $\log (P^{\rm SN}/{\rm day})=-1.1296$.

(iii) {\em Case 3} (see Fig. 3): The binary system is in the
optically thick wind phase at the moment of the SN explosion. The
binary in this case is ($M_2^{\rm i}$, $M_{\rm WD}^{\rm i}$, $\log
(P^{\rm i}/{\rm day})$) $=$ (1.80, 1.10, -0.60). Since the initial
mass of the WD is more massive than that of the other two cases
above, the WD may grow in mass to the Ch mass more easily, that is,
it is still in the optically thick wind phase when $M_{\rm
WD}=1.378\,M_{\odot}$, at which $M^{\rm SN}_2=1.2047\,M_{\odot}$ and
$\log (P^{\rm SN}/{\rm day})=-0.6232$.

(iv) {\em Case 4} (see Fig. 4): This case differs from the three
above. Before the SN explosion, the binary system experiences two
mass transfer phases. The binary in this case is ($M_2^{\rm i}$,
$M_{\rm WD}^{\rm i}$, $\log (P^{\rm i}/{\rm day})$) $=$ (1.25, 1.10,
-1.30). Due to the short initial orbital period of the system,
angular momentum loss induced by GWR leads the He star to fill its
Roche lobe before the exhaustion of central He, resulting in Case BA
mass transfer. The mass transfer continues to proceed until the mass
donor starts to shrink below its Roche lobe, terminating the mass
transfer phase. After it has exhausted He in its core, the He star
expands and fills its Roche lobe again when it evolves to the
subgiant stage, leading to Case BB evolution. The following
evolution of this system is similar to {\em Case 1}. When the
accreting WD grows to the Ch mass, the binary is in the He-shell
flash phase, and the binary parameters are $M^{\rm
SN}_2=0.6983\,M_{\odot}$ and $\log (P^{\rm SN}/{\rm day})=-1.1848$.

\subsection{Initial parameters for SN Ia progenitors}


\begin{figure*}
\epsfig{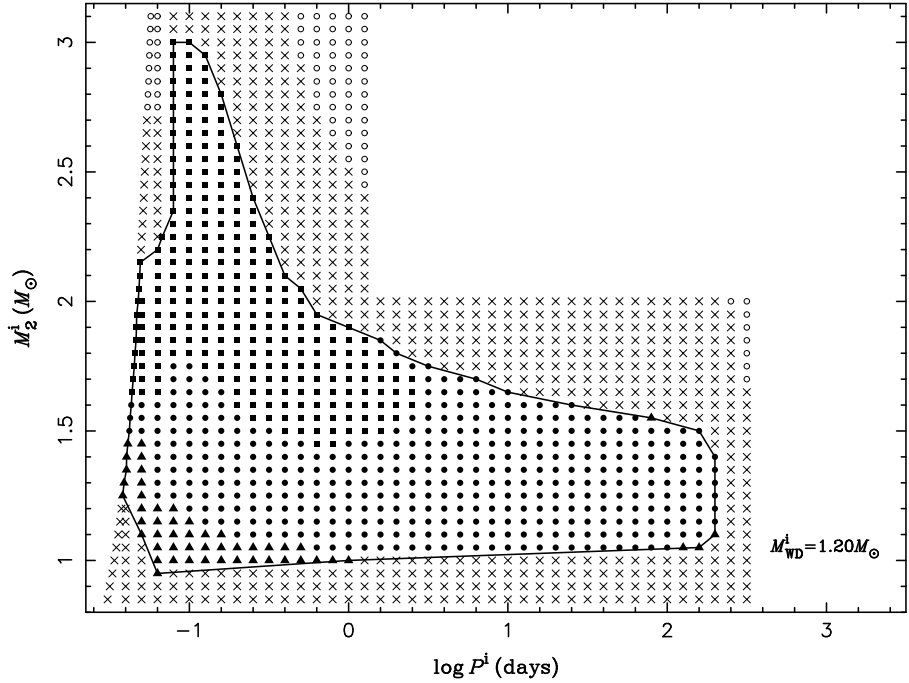} \caption{Final outcomes
of the binary evolution calculations in the initial orbital
period--secondary mass ($\log P^{\rm i}$, $M^{\rm i}_2$) plane of
the CO WD + He star system for initial WD mass of 1.2\,$M_{\odot}$.
The filled symbols are for those resulting in SN Ia explosions, that
is, the filled squares, circles and triangles denote that the WD
explodes in the optically thick wind phase, in the stable He-shell
burning phase and in the weak He-shell flash phase, respectively
(see {\em Cases 1-3} in Section 3.1). Crosses indicate systems that
may experience He novae, preventing the WD from reaching
$1.378\,M_{\odot}$, and open circles are those under dynamically
unstable mass transfer.} \label{20v}
\end{figure*}

\begin{figure*}
\epsfig{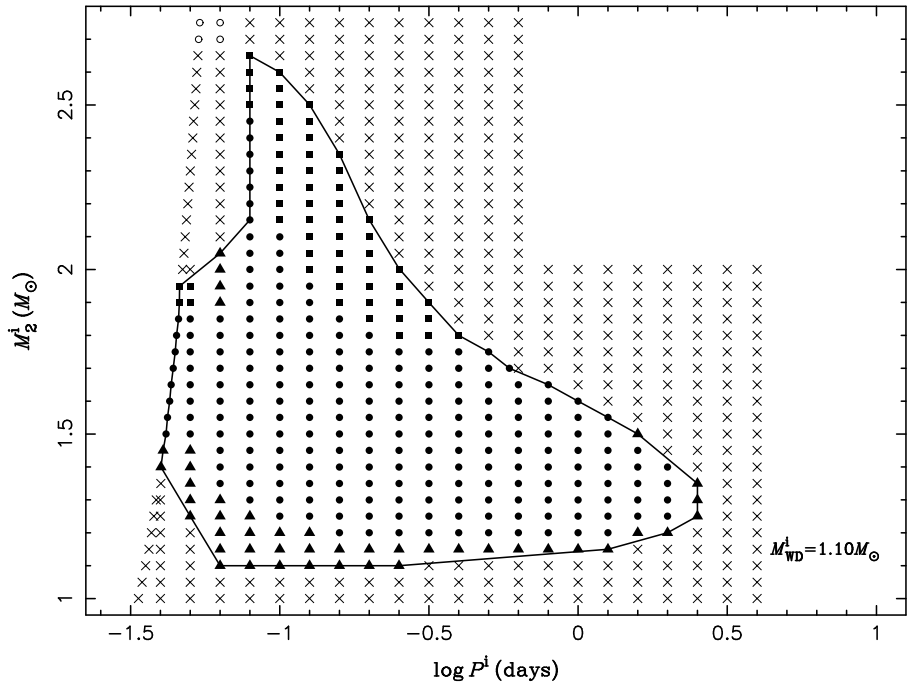} \caption{Similar to Fig.
5, but for initial WD mass of 1.1\,$M_{\odot}$.}
\end{figure*}

\begin{figure*}
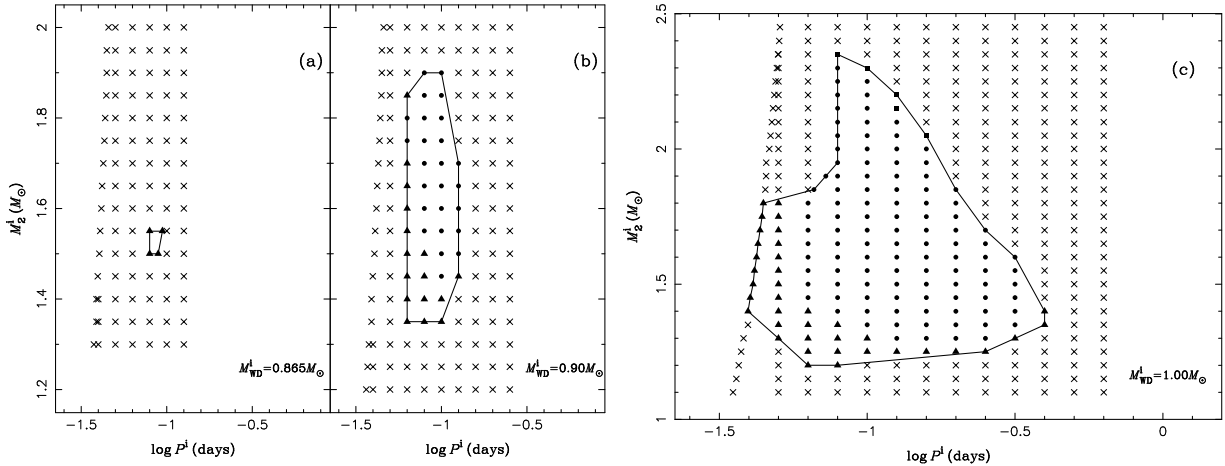

\centerline{\epsfig{file=fig2ab.ps,angle=270,width=8cm}\ \
\epsfig{file=fig2c.ps,angle=270,width=8cm}} \caption{Similar to Fig.
5, but for initial WD masses of 0.865, 0.9 and 1.0\,$M_{\odot}$.}
\end{figure*}

%

Figs 5-7 show the final outcomes of about 2600 binary evolution
calculations in the initial orbital period--secondary mass ($\log
P^{\rm i}-M^{\rm i}_2$) plane, where the filled symbols are for
those resulting in SN Ia explosions. The different cases described
in Section 3.1 are plotted with different symbols, that is, the
filled squares, circles and triangles denote that the WD explodes in
the optically thick wind phase, in the stable He-shell burning phase
and in the weak He-shell flash phase, respectively. Some WD + He
star systems fail to produce SNe Ia because of He nova explosions
(which prevents the WD growing in mass; the crosses in these
figures) or dynamically unstable mass transfer (resulting in a
common envelope; open circles in these figures). Note that, the left
boundaries of the initial $\log P^{\rm i}-M^{\rm i}_2$ plane in
these figures are determined by the minimum value of $\log P^{\rm
i}$, at which the He ZAMS star would fill its Roche lobe.


The contours of initial parameters for producing SNe Ia are also
presented in these figures. The left boundaries of the contours in
these figures (Figs 5-6 and panel c of Fig. 7) are set by the
condition that RLOF starts when the secondary is on the He ZAMS
\footnote{Note that, the upper parts of the left boundaries are
constrained mainly by a high mass transfer rate because of orbit
decay induced by GWR and large mass ratio, which leads to much of
the masses lost from the systems in a way of the optically thick
wind.}, while systems beyond the right boundary experience mass
transfer at a very high rate due to the rapid expansion of the He
stars in the subgiant stage and they lose too much mass via the
optically thick wind, preventing the WDs increasing their masses to
the Ch mass. The upper boundaries are also set mainly by a high mass
transfer rate but owing to a large mass ratio. The lower boundaries
are constrained by the facts that the mass transfer rate
$\dot{M}_{\rm 2}$ should be high enough to ensure that the WD can
grow in mass, and that the donor is sufficiently massive so that
enough mass can be transferred onto the WD to reach the Ch mass.



\begin{figure}
\epsfig{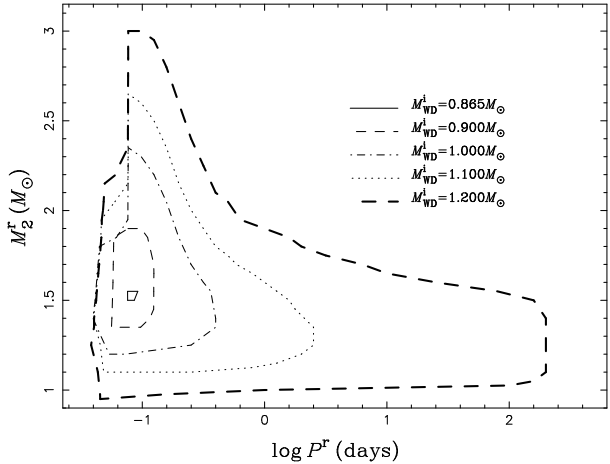} \caption{Regions in
orbital period--secondary mass plane ($\log P^{\rm r}$, $M^{\rm
r}_2$) for WD binaries (at the onset of the RLOF stage) that produce
SNe Ia for initial WD masses of $0.865, 0.90, 1.0, 1.1$ and
$1.2$\,$M_{\odot}$. The region almost vanishes for $M_{\rm WD}^{\rm
i}$=0.865\,$M_{\odot}$. } \label{contour}
\end{figure}


There is a time delay (up to about $10^{6}$\,yr) from the formation
of most WD + He star systems to the onset of RLOFs, and this time
delay varies with component masses and initial orbital periods. Fig.
8 shows the contours at the onset of RLOF for producing SNe Ia in
the $\log P^{\rm r}-M^{\rm r}_2$ plane for various WD masses (i.e.
$M_{\rm WD}^{\rm i}$ = 0.865, 0.90, 1.0, 1.1 and $1.2\,M_{\odot}$),
where $P^{\rm r}$ and $M^{\rm r}_2$ are the orbital period and the
mass of the He companion star at the onset of RLOF, respectively.
Note that, the enclosed region almost vanishes for $M_{\rm WD}^{\rm
i}=0.865\,M_{\odot}$, which is then assumed to be the minimum WD
mass for producing SNe Ia from this channel. If the parameters of a
CO WD + He star system at the onset of RLOF are located in the
contours, a SN Ia is then assumed to be produced. Thus, these
contours can be expediently used in BPS studies. One can send a
request to wangbo@ynao.ac.cn for the data points of these contours
and the interpolation FORTRAN code for these contours.

Note, the contours for SNe Ia between Figs 5-7 and Fig. 8 have some
differences, that is, Figs 5-7 are for initial parameters while Fig.
8 for those at the onset of RLOF. GWR may change the orbital period
during this time, especially for binaries with short orbital periods
(i.e. less than 1 day in this paper). For example, the left boundary
of the contour for $M_{\rm WD}^{\rm i}=1.2\,M_{\odot}$ in Fig. 8 has
shorter periods than that in Figs 5-7, because angular momentum loss
via GWR is more in short period binary systems than that in long
period systems.




\section{Birthrate of SNe Ia}
Adopting a prescription similar to that of Hachisu et al. (1999b)
and based on Fig. 8 of this paper, we can roughly estimate Galactic
SN Ia birthrate from the He star donor channel by using equation (1)
of Iben \& Tutukov (1984), i.e.
\begin{equation}
\nu = 0.2\,\Delta q \int_{M_{\rm A}}^{M_{\rm B}} {{d M} \over
M^{2.5}} \Delta \log A \, \mbox{yr}^{-1},
\label{realization_frequence}
\end{equation}
where $\Delta q$, $\Delta \log A$, $M_{\rm A}$ and $M_{\rm B}$ are
the appropriate ranges of the initial mass ratio and the initial
separation and the lower and upper limits of the primary mass for
producing SNe Ia in solar masses, respectively. We give the details
of the calculations in the following.

To estimate the birthrate of SNe Ia, we divide the initial WD mass
of $M_{\rm WD}^{\rm i}$ into three intervals, i.e. $0.9-1.0
M_\odot$, $1.0-1.1 M_\odot$, and $1.1-1.2 M_\odot$ (see also Hachisu
et al. 1999b). We ignore the range of $M_{\rm WD}^{\rm
i}=0.865-0.9\,M_\odot$, since its birthrate is too small to
contribute to the SN Ia birthrate as seen in Fig. 8. We use equation
(11) of Yungelson et al. (1995) to estimate $M_{\rm A}$ and $M_{\rm
B}$, i.e. the final core mass (CO WD mass) versus the ZAMS mass
relation,
\begin{equation}
\log {{M_{\rm WD}} \over {M_\odot}}= -0.22 + 0.36 \left( \log
{{M_{\rm i}} \over {M_\odot}}\right)^{2.5},
\end{equation}
where $M_{\rm WD}$ and $M_{\rm i}$ are the mass of the final CO WD
and of the ZAMS in solar masses, respectively. In addition, similar
to Hachisu et al. (1999b), we also use an approximation relation to
obtain $\Delta \log A$, i.e.
\begin{equation}
\Delta \log A \approx {2 \over 3} \Delta \log P^{\rm r},
\end{equation}
where $\Delta \log P^{\rm r}$ is taken from the SN Ia region in Fig.
8 and the factor of $2/3$ comes from the conversion between the
period and the separation. Take WD mass interval of
$1.1-1.2\,M_\odot$ as an example, we find that SN Ia explosions
occur for the ranges of $M_{\rm 1,i}=7.58-8.48\,M_\odot$, $M_{\rm
He,i}=0.95-3.0\,M_\odot$ and $\Delta \log A = 3.7 \times 2/3$, where
$M_{\rm 1,i}$ is the initial mass of the primary at ZAMS, and
$M_{\rm He,i}$ is the initial mass of the He star. Using Eggleton's
stellar evolution code, we can obtain the initial ZAMS mass $M_{\rm
i}$ = 5.60\,$M_{\odot}$ for the final He-core mass of
0.95\,$M_{\odot}$ (i.e. He star mass). Considering the initial mass
of the secondary at ZAMS is not more than that of the primary at
ZAMS, we constrain the range of $M_{\rm 2,i}$ from $5.60\,M_\odot$
to $M_{\rm 1,i}$, where $M_{\rm 2,i}$ is the initial mass of the
secondary at ZAMS. Thus, we obtain $\nu_{{\rm WD},1.1-1.2} \sim
0.82\times10^{-3}\,{\rm yr}^{-1}$ by substituting $\Delta \log A =
3.7 \times 2/3$, $\Delta q = 1-(5.60/8.48)=0.34$, $M_{\rm A} = 7.58$
and $M_{\rm B} = 8.48$ into equation (5). The SN Ia birthrates for
different WD mass intervals are summarized in Table 1. Note, for
different initial WD masses, the birthrate tendency of this channel
is different from that of WD + MS channel, that is, massive WDs for
producing SNe Ia have a high birthrate for the WD + He star channel.
This is because that massive WDs easily have massive He stars as
their companions.


Finally, the sum of SN Ia birthrates for three intervals
($0.9-1.0\,M_\odot$, $1.0-1.1\,M_\odot$, and $1.1-1.2\,M_\odot$)
gives $\nu\sim 1.2\times10^{-3}\,{\rm yr}^{-1}$ in the Galaxy, which
is lower than that inferred observationally (i.e. $3 - 4\times
10^{-3}\ {\rm yr}^{-1}$; van den Bergh \& Tammann 1991; Cappellaro
\& Turatto 1997).

\begin{table}
 \begin{minipage}{85mm}
 \caption{The birthrates of SNe Ia for different
WD mass intervals.}
 \label{tab5}
   \begin{tabular}{cccccc}
\hline \hline
$M_{\rm WD}^{\rm i}$ & $\Delta \log A$ & $M_{\rm A}$ & $M_{\rm B}$ & $\Delta q$ & $\nu_{\rm WD}$\\
$(M_\odot)$ & & $(M_\odot)$ & $(M_\odot)$ && ($10^{-3}$\,yr$^{-1}$)\\
\hline
$0.9-1.0$ & $1.0 \times 2/3$ & $5.60$ & $6.63$ & $0.02$ & $0.03$\\
$1.0-1.1$ & $1.8 \times 2/3$ & $6.63$ & $7.58$ & $0.18$ & $0.31$\\
$1.1-1.2$ & $3.7 \times 2/3$ & $7.58$ & $8.48$ & $0.34$ & $0.82$\\
\hline \label{1}
\end{tabular}
\end{minipage}
\end{table}

\section{DISCUSSION}
In our work, we have not considered the influence of rotation on the
He accreting WDs. The calculations of Yoon et al. (2004) showed
that, if rotation is taken into account, He burning is much less
violent than that without rotating. This may significantly increase
the accretion efficiency ($\eta _{\rm He}$ in this paper).
Meanwhile, the maximum stable mass of a rotating WD may be above the
standard Ch mass (i.e. the super-Ch mass model; Uenishi et al. 2003;
Yoon \& Langer 2005). However, in this paper we mainly focus on the
Ch mass explosions of the accreting WDs.

Yoon \& Langer (2003) (YL03) only considered a case BB evolution of
a CO WD + He star system which could produce a SN Ia. In this paper,
we systematically studed the He star donor channel for the
progenitors of SNe Ia (including case BA and case BB binary
evolution) and showed the contours of SNe Ia for various initial WD
masses. The binary studied in YL03 is located in the contours of
this paper. However, there are some differences between our
assumptions and theirs in mass accumulation efficiency. This is
mainly because that the effect of wind mass loss in YL03's
calculation is taken into account based an empirical formula of
Wolf-Rayet star mass loss, while we used an optically thick wind
model (Kato \& Hachisu 1994; Hachisu et al. 1996) and adopted the
prescription of KH04 for the mass accumulation efficiency of the
He-shell flashes onto the WDs.

The estimated birthrate of this channel is $\sim
1.2\times10^{-3}\,{\rm yr}^{-1}$, which is comparable to that of WD
+ MS channel (this is considered as an important channel to SNe Ia;
Han $\&$ Podsiadlowski 2004). Thus, the He star donor channel should
not be ignored when we study the progenitors of SNe Ia, and this
channel may increase the contribution of SD Ch scenario to SN Ia
birthrate in theory. Note, the estimated birthrate may be higher
than that in a real condition, since the long orbit period (i.e.
$\ga 1$ day in Fig. 8) systems considered to produce SNe Ia in
equation (1) of Iben \& Tutukov (1984) may not contribute to SNe Ia.
Moreover, Umeda et al. (1999) concluded that the upper limit mass of
CO cores born in binaries is about $1.07\,M_\odot$. If this value is
adopted as the upper limit of the CO WD, the birthrate of SNe Ia
from this channel will decrease to be $\sim 0.2\times10^{-3}\,{\rm
yr}^{-1}$ in the Galaxy.

Some precursory studies (Iben \& Tutukov 1994; YL03) suggested that
WD + He star systems may appear as supersoft X-ray sources (SSSs)
before SN Ia explosions. The initial orbital periods of WD + He star
systems producing SNe Ia for this channel range from $\sim$ 1\,h to
$\sim$ 200\,days, which may explain SSSs with various orbital
periods. we can estimate the X-ray luminosity of a WD + He star
system by $L_{\rm X}\sim \epsilon_{\rm He} |\dot{M_2}|$, where
$\epsilon_{\rm He}=6\times 10^{17}\,{\rm erg\,g^{-1}}$ is the energy
generation rate owing to He burning. A WD + He star system has
luminosity around $10^{37}-10^{38}\,{\rm erg\,s^{-1}}$ when He
burning is stable on the surface of the WD, consistent with that of
observed from SSSs ($L_{\rm X}=10^{36}-10^{38}\,{\rm erg\,s^{-1}}$;
Kahabka \& van den Heuvel 1997). Note, strong He II $\lambda4686$
lines are prominent in the luminous SSSs (see Kahabka \& van den
Heuvel 1997). Thus, we emphasize that SN Ia progenitors in the He
star donor channel may appear as SSSs during the stable He-shell
burning phase {\em without winds} \footnote{Massive wind may absorb
soft X-rays, which then cannot be detectable (Hachisu et al. 1999b;
Kato \& Hachisu 2003).}. By multiplying the birthrate, $\sim
1.2\times10^{-3}\,{\rm yr}^{-1}$, with an average duration of the
SSS phase, $\sim$ several times $10^5$\,yr, we estimate the current
number of this type of progenitors as SSSs in the Galaxy to be from
a few to several hundred. However, we do not observe such number of
WD + He star systems in SSS phase. This may attribute to the
Galactic interstellar absorption of SSSs (Di Stefano \& Rappaport
1994; Hachisu et al. 2008a). Moreover, Di Stefano \& Rappaport
(1994) also suggested that circumstellar matter may play a role in
the obscuration of X-rays (see also Hachisu et al. 2008a).


Generally, a CO WD + He star system is produced from an
intermediate-mass binary, which can be found in stellar populations
with relatively recent star formation. The minimum He star mass for
producing SNe Ia from this channel is $0.95\,M_{\odot}$, with which
we can roughly estimate the maximum delay time of this channel. The
initial ZAMS mass $M_{\rm i}$ = 5.6\,$M_{\odot}$ can produce He star
mass of 0.95\,$M_{\odot}$ (see Sect. 4). The MS lifetime is about
81\,Myr for a star of 5.6\,$M_{\odot}$ (Eggleton 2006). Moreover, we
should consider the MS lifetime of He stars, since the He stars in
long orbital period systems experience RLOF after the exhaustion of
central He. The MS lifetime of $0.95\,M_{\odot}$ He star is about
18\,Myr. Thus, the maximum delay time of this channel is about
$10^{8}$\,yr, i.e. the He star donor channel can explain SNe Ia with
short delay times ($\la 10^{8}$\,yr; Mannucci et al. 2006; Aubourg
et al. 2008). Note, Hachisu et al. (2008a) investigated new
evolutionary models for SN Ia progenitors, introducing the
mass-stripping effect on a MS or slightly evolved companion star by
winds from a mass-accreting WD. The model can also explain the
presence of very young ($\la 10^{8}$\,yr) populations of SN Ia
progenitors (see also Hachisu et al. 2008b).

When WD + He star systems explode as SNe Ia, the SN Ia spectra may
more or less show sign of He. Unfortunately, He associated with SNe
Ia fail to be detected in the past years (Marion et al. 2003;
Mattila et al. 2005). Therefore, Kato et al. (2008) suggested that
SNe Ia from this channel may be rare. However, further study on this
channel is necessary, since this channel can explain some SNe Ia
with short delay times. In addition, GWR is strong in WD + He star
systems with short orbital periods. Thus, the WD + He star systems
with short orbital periods are possibly detectable sources of GWR.

\section{SUMMARY}
Using an optically thick wind model (Kato \& Hachisu 1994; Hachisu
et al. 1996) and adopting the prescription of KH04 for the mass
accumulation efficiency of the He-shell flashes onto the WDs, we
systematically studied the He star donor channel for producing SN Ia
progenitors and showed the parameter spaces for SNe Ia from this
channel. The study shows that this channel is noteworthy for
producing SNe Ia, that is, the birthrate of SNe Ia is $\sim
1.2\times 10^{-3}\ {\rm yr}^{-1}$ in the Galaxy. In addition, the
minimum mass of CO WD for producing SNe Ia from this channel may be
as low as $0.865\,M_{\odot}$. According to the orbital period range
and X-ray luminosities of WD + He star systems, we considered that
WD + He star systems may appear as SSSs before SN Ia explosions,
which is consistent with YL03. Importantly, the He star donor
channel can explain SNe Ia with short delay times ($\la
10^{8}$\,yr). Meanwhile, WD + He star systems with short orbital
periods may be detectable sources of GWR, and the contours of $\log
P^{\rm r}-M^{\rm r}_2$ plane at the beginning of RLOF can be used in
BPS studies. In future investigations, we will study the properties
of companions at the moment of SN explosion, which might be verified
by future observations.



\section*{Acknowledgments}
We thank an anonymous referee for his/her valuable comments which
help to improve the paper. Bo Wang thanks Prof. M. Kato of Keio
University for helpful discussions. Bo Wang also thanks prof. Ph.
Podsiadlowski of Oxford University for helpful discussions during
his visit to NAOC/YNAO in April 2008. This work is supported by the
National Natural Science Foundation of China (Grant Nos. 10433030,
10521001, 2007CB815406 and 10603013) and the Foundation of the
Chinese Academy of Sciences (Grant No. 06YQ011001).

\label{lastpage}
\end{document}